\documentclass[12pt]{iopart}

\usepackage[
    backend=biber,
    style=numeric-comp,
    sortlocale=en_US,
    natbib=true,
    url=false, 
    doi=true,
    eprint=false,
    sorting=none,
    giveninits=true,
    maxnames=10,
    minnames=10,
    alldates=year
]{biblatex}
\renewbibmacro{in:}{}

\usepackage{hyperref}
\usepackage{braket}
\usepackage{graphicx}
\usepackage[noabbrev,nameinlink]{cleveref}
\usepackage{xcolor}
\usepackage{subfig}
\usepackage{parskip}        

\usepackage[labelfont=bf]{caption}
\DeclareCaptionLabelFormat{rightparen}{ \fontfamily{phv}\selectfont {{\textbf{#2)}}}}
\begin{document}

\title[]{A tunable quantum random number generator based on a fiber-optical Sagnac interferometer}

\author{Joakim Argillander$^1$, Alvaro Alarc\'{o}n$^1$ and Guilherme B. Xavier$^1$}
\address{$^1$ Institutionen f\"{o}r Systemteknik, Link\"{o}ping University, 581 83 Link\"{o}ping, Sweden}
\ead{joakim.argillander@liu.se}
\vspace{100pt}

\vspace{10pt}
\begin{indented}
\item[]January 2022
\end{indented}

\begin{abstract}
Quantum random number generators (QRNG) are based on the naturally random measurement results performed on individual quantum systems. Here, we demonstrate a branching-path photonic QRNG implemented with a Sagnac interferometer with a tunable splitting ratio. The fine-tuning of the splitting ratio allows us to maximize the entropy of the generated  sequence of  random numbers and effectively compensate for tolerances in the components. By producing single-photons from attenuated telecom laser pulses, and employing commercially-available components we are able to generate a sequence of more than 2 gigabytes of random numbers with an average entropy of 7.99 bits/byte directly from the raw measured data. Furthermore, our sequence passes randomness tests from both the NIST and Dieharder statistical test suites, thus certifying its randomness. Our scheme shows an alternative design of QRNGs based on the dynamic adjustment of the uniformity of the produced random sequence, which is relevant for the construction of modern generators that rely on independent real-time testing of its performance.
\end{abstract}

%
\noindent{\it Keywords}: quantum random number generation, tunable beamsplitter, fiber-optic Sagnac interferometer

%
%
%
%
\newpage

\section{Introduction}
Quantum information science is an interdisciplinary field where computer science meets physics, and where long lasting principles of information transfer, cryptography and secret sharing are challenged \cite{NielsenBook}.  Quantum random number generators (QRNG) are mature devices that directly benefit from the properties of  quantum systems in order to produce truly random numbers \cite{Herrero-Collantes2017}. By harnessing the fundamental randomness from the outputs of quantum mechanical measurements it is possible to generate sequences of fundamentally unpredictable random numbers. A popular implementation for QRNGs is based on performing measurements on single-photons. Typically, in these photonic QRNGs, one degree of freedom of a photon is used to prepare a superposition state of the form $\ket{\psi} = \alpha\ket{0} + \beta\ket{1}$, where $\left|\alpha\right|^2 + \left|\beta\right|^2 = 1$, which is then projected onto the basis formed by $\ket{0}$ and $\ket{1}$. By letting one of the basis vectors represent a binary ``0'' and the other a binary ``1'', we can utilize the fact that measurements of an equally weighted superposition state yield uniformly distributed random measurement outcomes to generate a random sequence by performing repeated measurements. 

One type of photonic QRNG is known as a branching-path generator \cite{Herrero-Collantes2017, Jennewein2020, Stefanov2000, Wang2006, Naruse2015}, and is typically implemented either using polarized light and polarizing beamsplitters, or with unpolarized light and conventional beamsplitters. Regardless of the scheme, it is of utmost importance that the beamsplitters' splitting ratio is $50:50$ for transmitted and reflected light respectively, in order to guarantee an unbiased sequence. In practice, due to manufacturing limitations, the splitting ratio will not be perfectly balanced, and this leads to non-uniformity of the bit distribution of the generated sequence. Furthermore, the splitting ratio may also change over time, due to temperature fluctuations and other environmental factors, as well as component degradation. Branching-path generators may also suffer from detector imbalances, where two identical detectors used at each of the beamsplitter's outputs have slightly different detection efficiencies that also lead to a bias in the sequence \cite{Herrero-Collantes2017}. Such a bias reduces the unpredictability of the sequence, and also so does its usability, not only for cryptography, but also for uses like Monte-Carlo simulations and gambling. For cryptography specifically, the security is directly proportional to the randomness of the secret key, and therefore a low-entropy sequence increases the probability that an attacker is able to successfully guess the key, which thus reduces the effective strength of the cipher \cite{Goldwasser1984}.

In this article, we demonstrate a tunable-ratio branching-path photonic QRNG by implementing a dynamically tunable beamsplitter using a fiber-optical Sagnac interferometer. The tunable beamsplitting function is implemented with an active phase modulator placed in a Sagnac loop working as a single-photon router \cite{Alarcon2020}. By modulating the phase of a light packet propagating in the clockwise direction vs the counter-clockwise one, we are able to tune the output probabilities from the Sagnac interferometer through a change in the amplitude of the driving signal fed to the phase modulator. This allows us to change the entropy of the generated sequence as we wish. Both outputs are then detected through a time-division multiplexing scheme to allow detection with only one single-photon detector. {\color{black}Interferometers in a Sagnac-type configuration have previously been used to modulate polarization \cite{Agnesi2019,Avesani2021} and intensity \cite{Roberts2018} in order to prepare states for quantum key distribution experiments. While building on the same principle, our system is the first demonstration of interferometric routing of single photons for quantum random number generation.} In this work we are able to generate a continuous stream of random numbers over 36 hours of continuous operation at an average rate of 131 kbit/s. The generation rate and entropy are highly stable over time due to the intrinsic phase stability of the Sagnac interferometer configuration. We finally demonstrate the randomness of the sequence by passing it successfully through the stringent statistical test suites NIST 800-22 \cite{Rukhin2010} and Dieharder \cite{R.Brown2011} aimed at assessing random number generators. A critical issue in QRNGs, especially the ones aimed at cryptographic applications, is whether the user can trust the device. A solution to this problem is performing independent real-time testing of the internal devices of the QRNG, usually through the user changing either the prepared quantum states, or a measurement parameter \cite{Ma2016}. {\color{black} In order to test the device and achieve high confidence that the generated sequence is private, a high interferometric visibility is necessary \cite{Carine2020}}. The dynamic tunability provided by our setup is therefore useful to tackle this issue, and is particularly appealing to recently demonstrated measurement-device-independent QRNGs \cite{Nie2016, Carine2020}. Our results show that Sagnac interferometers can be successfully employed as tunable beamsplitters for the construction of quantum random number generators, thus opening up for implementation of state-of-the-art measurement-device independent quantum random number generators.

\section{The experiment}

\begin{figure}[t!]
    \centering
    \includegraphics[clip, trim=0cm 3.7cm 0cm 0cm, width=0.4\textwidth, angle=0]{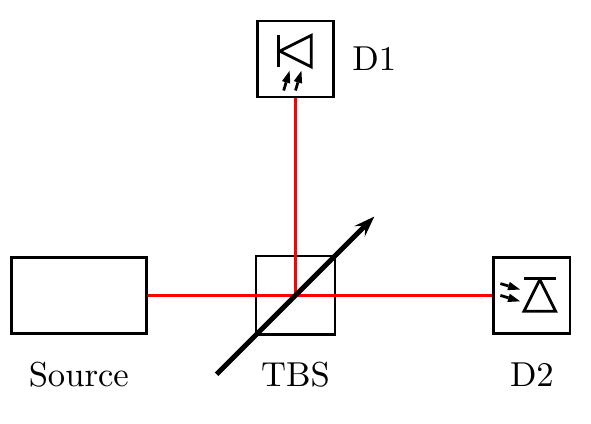}
    \caption{\label{fig:simpleexpsetup} Principle of a tunable-ratio beamsplitter for quantum random number generation. Light from a photon source is split using a tunable beamsplitter (TBS). Detectors (D1, D2) placed at the outputs of the TBS register photon events and record a binary ``0'' or a binary ``1'' depending on whether the photon impinges on D1 or D2. The choice of assignment of ``0''/``1'' to D1/D2 is arbitrary.}
\end{figure}

{\color{black}The principle for a QRNG scheme using a TBS is based on a source of strongly attenuated pulses, a beamsplitter with a variable splitting ratio and single-photon detection of the two outputs of the beamsplitter (\cref{fig:simpleexpsetup}).} A detection at one output is assigned the bit ``0'', while ``1'' is assigned to the other one. The entire system works synchronously, driven by Field Programmable Gate Array (FPGA) electronics, which also performs the data acquisition. {\color{black} If both outputs are simultaneously detected (either from two photons generated together, or from noise sources), the FPGA discards that result in order to remove the effects of after-pulsing and multiphoton events}. Similarly, if no photons are detected for a generated pulse (due to absorption) no result is recorded. The generated bits are then streamed from the FPGA to a personal computer for randomness extraction and storage. The bitstream then passes through statistical tests to certify its randomness.

\begin{figure}[t!]
    \centering
    \includegraphics[clip, trim=0cm 3cm 0cm 0cm, width=\textwidth, angle=0]{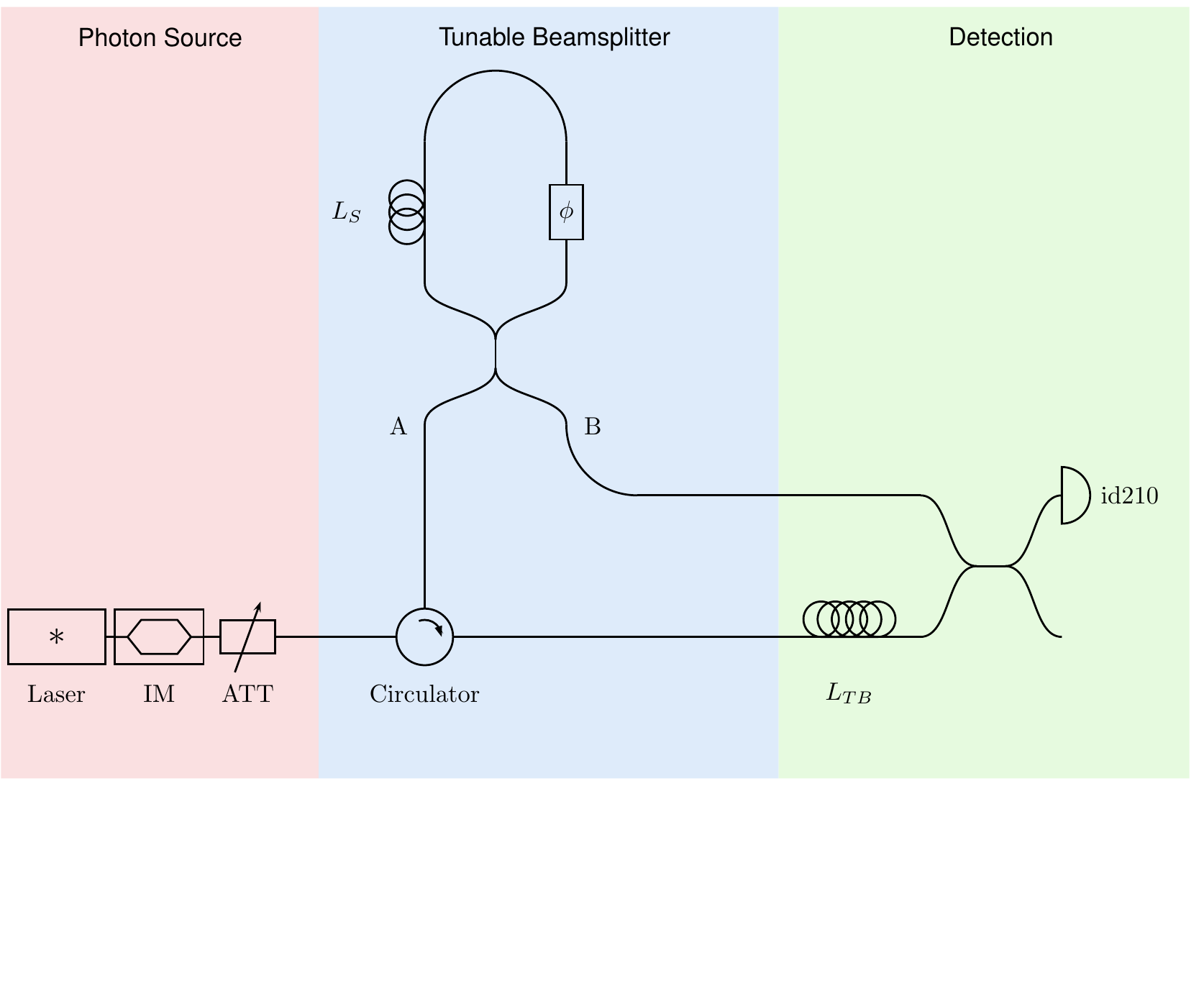}
    \caption{\label{fig:expsetup}Experimental setup. The CW-laser is first fed to an intensity modulator (IM) that chops the laser into 20 ns pulses, which are then attenuated using a variable attenuator (ATT) to weak coherent pulses (WCP) with an average photon probability of approximately $10$ photons per pulse. The WCPs are then fed to a Sagnac interferometer that acts as a tunable beamsplitter where the superposition $\ket{\psi}= 1/\sqrt{2}\left(\ket{\textrm{CW}} + ie^{i\phi}\ket{\textrm{CCW}} \right)$ is prepared, and CW and CCW denote the photon state of propagation inside the Sagnac loop. At the output of the Sagnac loop, we use a {\color{black} 150 meter} fiber-optic delay line ($L_{TB}$) to add {\color{black}$\approx 750$} ns of delay to one term, effectively time-bin multiplexing the two outputs from the Sagnac interferometer into an \emph{early} and a \emph{late} time-bin for detection using one single-photon detector.}
\end{figure}

The key aspect in our experiment is the use of a tunable beamsplitter (\cref{fig:simpleexpsetup}). Although in principle it is possible to create one by adding extra attenuators in each output of a conventional beamsplitter, or by using a half-wave plate in tandem with a polarizing beamsplitter, these options do not provide possibilities of fast tuning. Furthermore, in the case of the first option, extra losses are added to the system. Here we pursue an interferometric approach to implement tunability. For instance, a Mach-Zehnder interferometer can realize a tunable beamsplitter, with the tunability provided by setting the relative phase $\phi$ between the interferometer arms \cite{Ma2011}. The output probability of the single-photon is then given by a sinusoidal function of $\phi$. However, the drawback of using conventional Mach-Zehnder interferometers is that they suffer from intrinsic phase instability, and therefore require either mechanical isolation (i.e. dampened tables, environmental protection, etc) \cite{Micuda2014} or active stabilization \cite{Xavier2011,Reynaud1992}. In our work we follow the approach of \cite{Alarcon2020} where we employ a Sagnac interferometer to provide the same functionality, with the added benefit of intrinsic phase stability. In our Sagnac interferometer a high-speed telecom fiber-pigtailed phase modulator is placed to give a relative phase shift $\phi$ between the two propagation directions (\cref{fig:expsetup}). This is done by synchronizing the phase shift with the wave packet, such the phase shift is only applied in one direction by the modulator. 

We employ a continuous wave (CW) distributed feedback telecom laser diode, centered at $\lambda$ = 1546.92 nm, connected to variable optical attenuators (ATT) to produce weak coherent states. {\color{black}The FPGA board generates electrical pulses at a repetition rate of $250$ kHz in order to drive an intensity modulator which chops the CW input light into optical pulses.} The electronic pulses are shaped to $20$ ns width, as limited by the employed electronics driver before being fed to the intensity modulator. The attenuated optical pulses pass through an optical circulator before entering one of the input ports of a bidirectional $50:50$ fiber coupler. The output ports form the Sagnac loop, together with a phase modulator and a 50 m long optical fiber spool ($L_s$) that works as a delay line (\cref{fig:expsetup}). The delay line is needed to ensure there is sufficient time separation for the phase modulation signal to act on the wave packet propagating in only in one direction and not in the opposing one, thus creating the necessary relative phase shift to generate the different output probabilities in the interferometer outputs \cite{Alarcon2020}. {\color{black}Not shown in \cref{fig:expsetup} are two manual polarization controllers added in the Sagnac loop, where one is used to align the pulses' polarization with the input polarization of the phase modulator, while the second one is used to ensure that the two counter-propagating terms have the same polarization at the recombining beamsplitter.}

Inside the Sagnac loop the attenuated pulses can be represented as a coherent superposition of the two counter-propagating paths as $\ket{\psi} = 1/\sqrt{2}(\ket{\textrm{CW}} + ie^{i\phi}\ket{\textrm{CCW}})$ where $\ket{\textrm{CW}}$ and $\ket{\textrm{CCW}}$ represent the clockwise and counter-clockwise propagation paths respectively. The detection probabilities at the outputs A and B of the Sagnac loop are proportional to $\cos^2(\frac{\phi}{2})$ and $\sin^2(\frac{\phi}{2})$. For voltages corresponding to $\phi =\pi/2, 3\pi/2$ the Sagnac works as a beamsplitter with a perfectly balanced splitting ratio. The phase shift $\phi$ is proportional to the modulator's driving voltage, thus allowing us to precisely control the splitting ratio between 0 and 1 in the ideal case. We fine-tune the desired ratio by applying a small offset $\delta V_\phi$ to the theoretical voltage value $V_\phi$. Therefore, the voltage we apply to achieve a balanced splitting ratio is $\overline{V}_{\pi/2} = V_{\pi/2} + \delta V_{\pi/2}$. {\color{black} Due to insertion losses and propagation losses in the time-bin multiplexer, a balanced splitting ratio between the early and late time bin does not imply a 50:50 probability for a photon to emerge at the outputs A:B of the Sagnac interferometer.} The two outputs of the Sagnac (denoted A and B in \cref{fig:expsetup}) are then routed to the detection system (with output A passing through the circulator to decouple it from the attenuated pulses coming from the source, propagating in to the Sagnac loop). We opted to use one SPD and mapped each output path to an \textit{early} or a \textit{late} time-bin, with both time slots detected sequentially by the SPD. The time-bin separation is created by a 750 ns time delay (a $\approx$150 m optical fiber spool) {\color{black}placed at the A output path} from the Sagnac loop. The two paths are then recombined with a 50:50 bidirectional fiber coupler, and one of its outputs is connected to the SPD (idQuantique id210), operating in Geiger mode with 10\% detection efficiency, 2.5 ns gate width and approximately $10^{-5}$ %
dark count probability per gate. The SPD is externally gated by the FPGA by sending pairs of pulses, each separated by the time-bin delay, that are synchronized with the signal sent to the intensity modulator that generates the photon pulses. {\color{black}Each of these pulses correspond to a gate in the \emph{early} and \emph{late} bin respectively.} The output signals from the SPD are also read by the FPGA for data acquisition. We set the deadtime of the SPD to 500 ns, so that it is shorter than the time separation between the \emph{early} and \emph{late} time-bins. We then let a detection of a photon in the \emph{early} time-bin represent a ``0'' and a detection in the \emph{late} time-bin represent a ``1''.

\section{Results}

\begin{figure}[t!]
    \centering

    \subfloat[]{\label{fig:countsphasevoltage}\includegraphics[width=.49\textwidth]{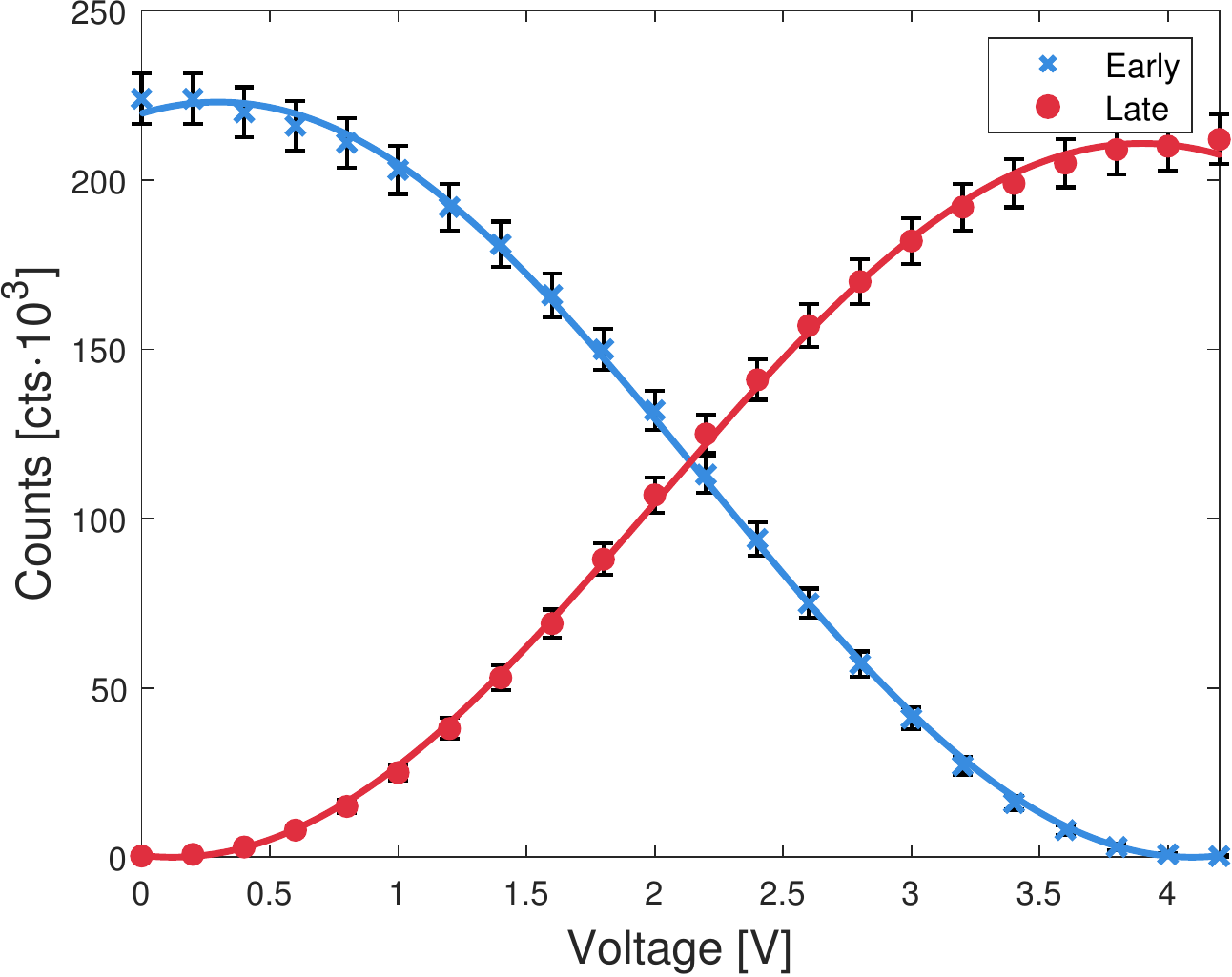}}
    \hspace{1em}
    \subfloat[]{\label{fig:entropysweep}\includegraphics[width=.47\textwidth]{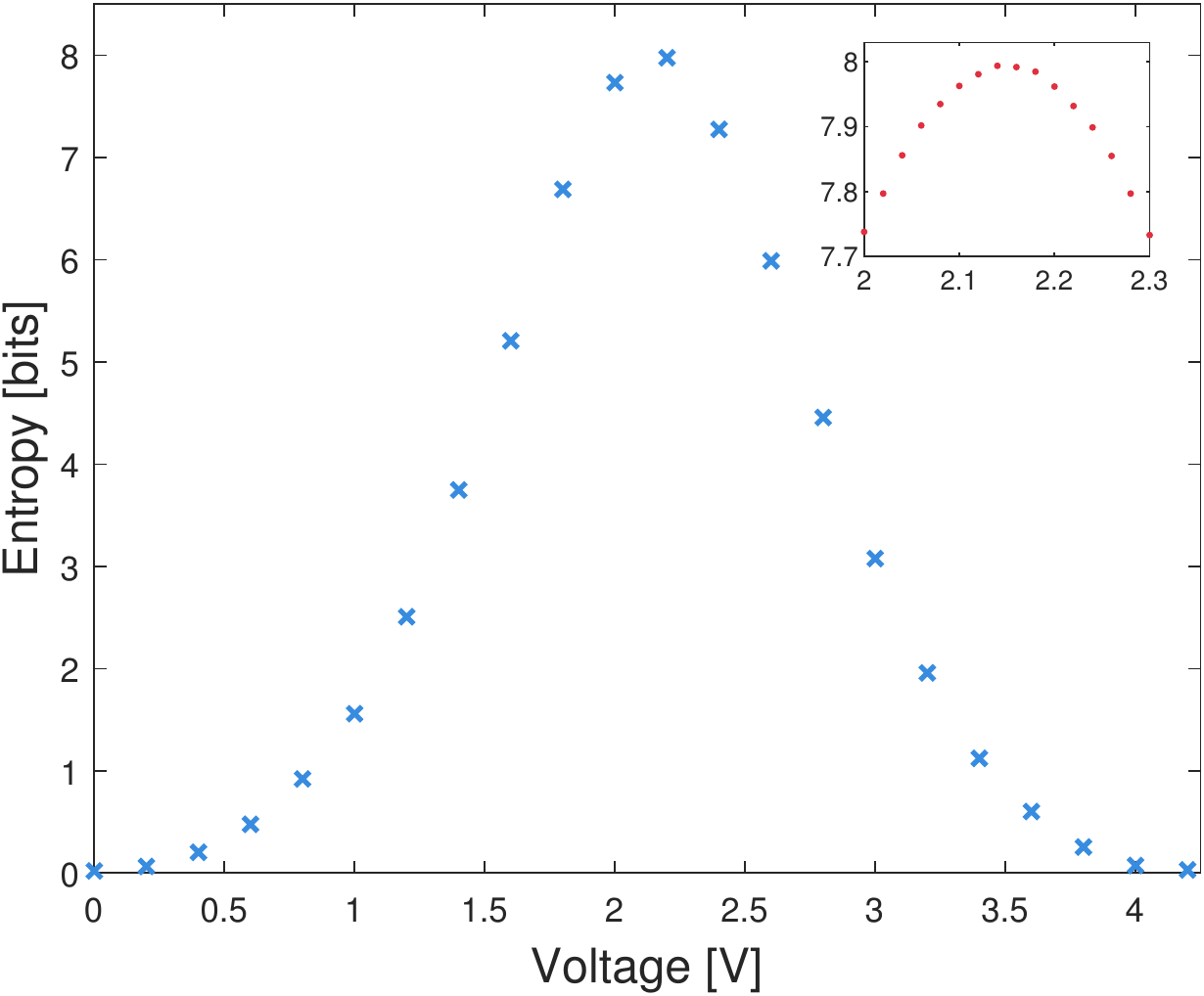}}

    \caption{\label{fig:extinctionentropy} a) Characterization of detected photon events measured in the \emph{early} and \emph{late} time-bin, as a function of applied voltage to the phase modulator. The error bars are calculated assuming Poissonian photon statistics ($\sigma = \sqrt{N}$ for a photon count of $N$). Also plotted is a fit of the theoretical detection probabilities that are proportional to $\cos^2$ [$\sin^2$] for the \emph{early} [\emph{late}] time-bin, respectively. b) Raw entropy (before extraction) as a function of applied voltage to the phase modulator Also shown in the inset is a plot of the entropy versus applied voltage to the phase modulator with a smaller step size. }
    
\end{figure}

First we demonstrate the variable splitting ratio of the Sagnac interferometer acting as a tunable beamsplitter. We set the attenuators such that we maximize the detection probability at the detector for both time-bins, while still operating in a linear regime (i.e. the detector is not saturating).
This is observed from \cref{fig:countsphasevoltage}, where the number of detection events per integration time fits well with the theoretical prediction that output A [B] from the Sagnac loop should follow a $\cos^2(\frac{\phi}{2})$ [$\sin^2(\frac{\phi}{2})$] distribution. We also observe from these results that the Sagnac interferometer is operating correctly, and complementary output behaviour of the two outputs is observed, as expected.  

We then wish to optimize the operating point of the Sagnac loop for optimal random number generation. We scan the driving voltage of the phase modulator and measure the Shannon entropy of the generated bits when considered as $8$-bit symbols. The Shannon entropy $\mathcal{H}$ is defined as $\mathcal{H} = -\sum_{i=0}^{255}p_i\log_2{(p_i)}$, where $p_i$ is the probability of finding the $i_{th}$ 8-bit sequence. We increase the voltage $V_\phi$ applied to the phase modulator in the range $0\leq V_\phi \leq 4.2$ in steps of $200$ mV and measure the entropy, as can be seen in \cref{fig:entropysweep}. An inset is shown in \cref{fig:entropysweep} where we also sweep the voltage between $2-2.3$ volts in steps of $20$ mV, and likewise measure $\mathcal{H}$ to identify the optimal operating point where the entropy is maximized. It can be clearly seen that the maximum obtained entropy (7.98 bits/byte) occurs for an applied driving voltage of around $2.15$ V.

\begin{figure}[t!]
    \centering
    \subfloat[]{\label{fig:entropytime}\includegraphics[clip, trim=0cm 5.1cm 0cm 0cm, width=.48\textwidth]{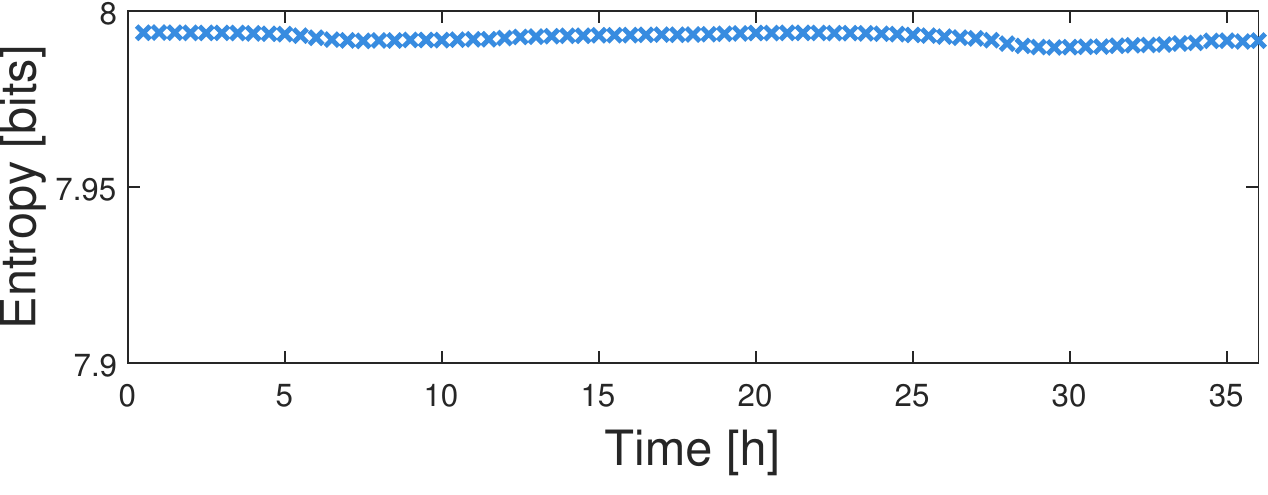}}
    \hfill
    \subfloat[]{\label{fig:bitratetime}\includegraphics[clip, trim=0cm 0cm 0cm 5.28cm, width=.48\textwidth]{figures/bitrate_entropy_over_time-eps-converted-to.pdf}}

    \caption{\label{fig:entropybitrate} Characterization of the stablity of the raw bitrate  a) and entropy stability b) over a period of 36 hours. Each of the data points represents the average over 30 minutes.}
\end{figure}

\begin{figure}[t!]
    \centering
    \subfloat[]{\label{fig:nisttable}\includegraphics[clip, trim=0.4cm 0cm 1.2cm 0.5cm, width=.5\textwidth]{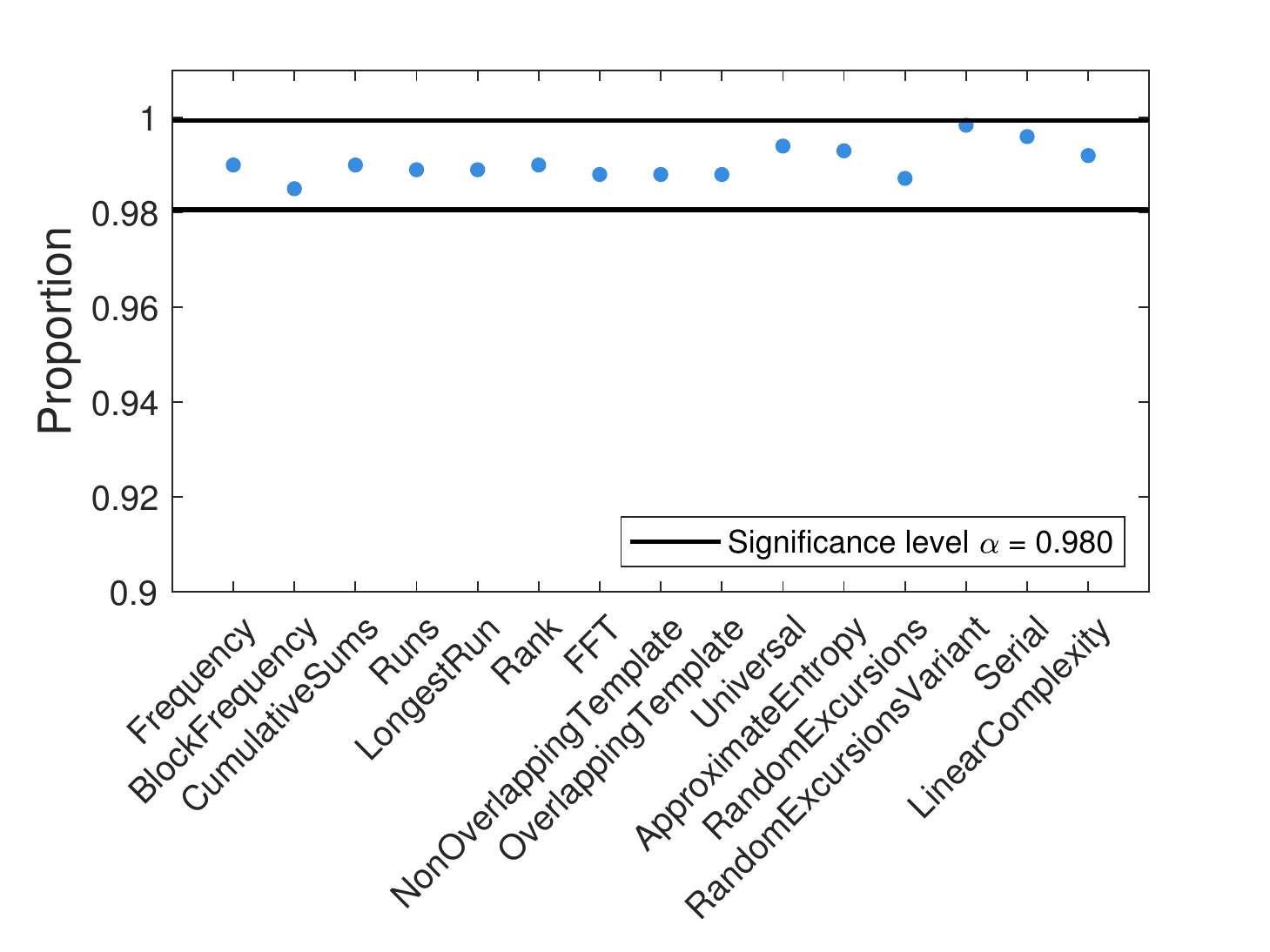}}
    \subfloat[]{\label{fig:diehardertable}\includegraphics[clip, trim=0cm 0cm 1.3cm 0.5cm, width=.5\textwidth]{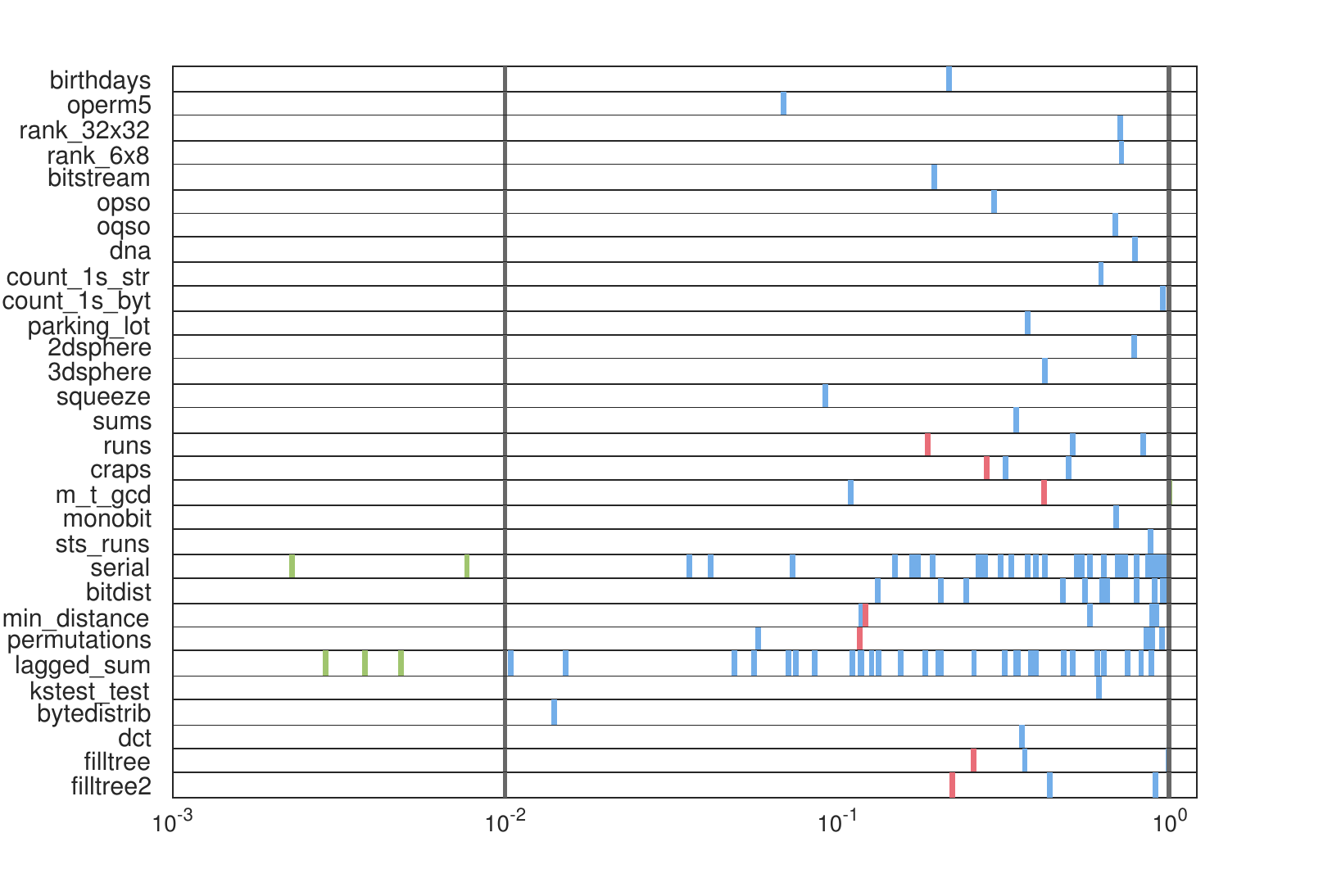}}

    \caption{\label{fig:statisticalresults} Results from statistical testing of randomness from NIST STS 800-22\cite{Rukhin2010}, a) and Dieharder\cite{R.Brown2011} b). In \cref{fig:nisttable} we plot the proportion of tests that have passed, along with the confidence interval that the proportion must fall within. For \cref{fig:diehardertable}, each colored vertical line corresponds to a p-value from the tests. Lines in green are p-values that fall outside of the confidence bound $10^{-2} < p < 10^0$. The red lines are the resulting p-value of a Kolmogorov-Smirnov test done on test results that yield multiple p-values. }
\end{figure}

With the entropy maximized, we then proceed to acquire a long continuous sequence of bits to feed to the statistical randomness tests, and also to demonstrate the robustness of our implementation. The raw bit sequence is buffered in the FPGA in blocks of 32 kilobytes and then transmitted over an Ethernet connection to a personal computer for storage. Before assessing the sequence, it is passed through a Toeplitz randomness extraction procedure in order to remove any remaining non-uniformities from the raw data. To extract randomness, we split the generated sequence into $N$ sequences of length $n$, and multiply each subsequence with an $n\times m $ binary Toeplitz matrix which yields an output of $m$ hashed bits \cite{Qi2017}. We choose to use $n=400$, and as defined in \cite{Qi2017}, we obtain $m = \mathcal{H}_\mathrm{min} -2\log_2\varepsilon$ where $\mathcal{H}_\mathrm{min}$ is the min-entropy of 8-bit strings which is defined as {\color{black}$\mathcal{H}_\mathrm{min} = -\log_2 \left[ \max_{x\in\left\{0, 1\right\}^8} \mathrm{Pr}\left\{X=x\right\}  \right] $. We measure a min-entropy of $7.7451$ from the acquired sequence before extraction.} We choose a security parameter $\varepsilon=2^{-100}$ which is derived from the leftover hash lemma \cite{Ma2013}. We seed the Toeplitz matrix $T$ at the beginning of the extraction procedure with $n+m-1$ bits taken from the raw sequence, as it is sufficient to only specify the first row and first column. The matrix $T$ can be reused for each of the subsequences, and we obtain the extracted sequence of bits by concatenating the results of the multiplications $Tn_1, Tn_2, Tn_3,...,T_N$, where the subscript represents the subsequence index \cite{Qi2017, Ma2013}. The extraction procedure is lossy, and for our choice of parameters, we are able to achieve an efficiency of $\approx 50\%$. {\color{black} After extraction we calculate the min-entropy of the sequence to be $7.9982$ bits.}

While recording the raw data, we also measure the entropy of each streamed 32 kilobyte block, and as can be seen in \cref{fig:entropytime}, we achieve an average entropy of $7.9925 \pm 0.0014$ bits/byte over $36$ hours, which demonstrates exceptional phase stability of the QRNG. Likewise, we also measure the raw bitrate when streaming the data, and we achieve an average bitrate of $131.8 \pm 2.3$ kbps (\cref{fig:bitratetime}). The bitrate is mainly limited by the delays in the electronics that amplify the signals from the FPGA in order to drive the phase modulator, and in the triggering electronics which is necessary to synchronize the generation of pulses with the detection. By improving the driver electronics, the bitrate is only limited by the separation of the time bins.

After generating $2.2$ gigabytes of data, and extracting it, 
we pass the sequence through the NIST 800-22 Statistical Test Suite\cite{Rukhin2010} (NIST test) and the Dieharder\cite{R.Brown2011} test suite. Both test suites subject the random sequence to a number of statistical tests aimed at assessing the randomness of any given sequence. We run the NIST tests with $1000$ bitstreams of length $1$ Mb. To deem any given test as passed, we form a confidence bound as defined in \cite{Rukhin2010} for the proportion of bitstreams that must yield a p-value  $p \geq 10^{-2}$ using the formula $\hat{p} \pm 3\sqrt{\frac{\alpha(1-\alpha)}{n}}$ where $\alpha=10^{-2}$, $n=1000$ is the number of bitstreams and $\hat{p}=1-\alpha$. This gives us a confidence interval of $0.980 \leq r < 0.999$, and if the ratio of tests that have passed falls outside this range, there is evidence that the sequence is not random \cite{Rukhin2010}. \Cref{fig:nisttable} shows the proportions of tests that pass, for any given test category, along with the significance level.

The Dieharder test suite yields a verdict ("Fail", "Weak", "Pass") for each test, and a test is deemed to pass if the p-value associated with it is in the interval $10^{-2} < p < 1-10^{-2}$. As can be seen in \cref{fig:diehardertable}, there are parameterized tests where each test yields multiple p-values (one for each value of the parameter). We plot all outputted p-values, and compute a resulting p-value from a Kolmogorov-Smirnov test of all the p-values. As is well-known, the p-values should be uniformly distributed under the null hypothesis that the sequence is indeed random, and the Kolmogorov-Smirnov test is used to eliminate false rejection of the null hypothesis (Type-I error) \cite{Hung1997}. 
For three of the tests, we observe that the resulting p-values fall outside the confidence bound, and for those values of the parameter, the test verdict is "weak". Since the Dieharder suite is a statistical test, it is probable that there will be occasional "weak" results. If the generator truly is weak, then it will certainly fail other tests \cite{R.Brown2011}.

\begin{figure}[t!]
    \centering

    \includegraphics[clip, trim=0cm 0cm 0cm 0cm, width=.6\textwidth]{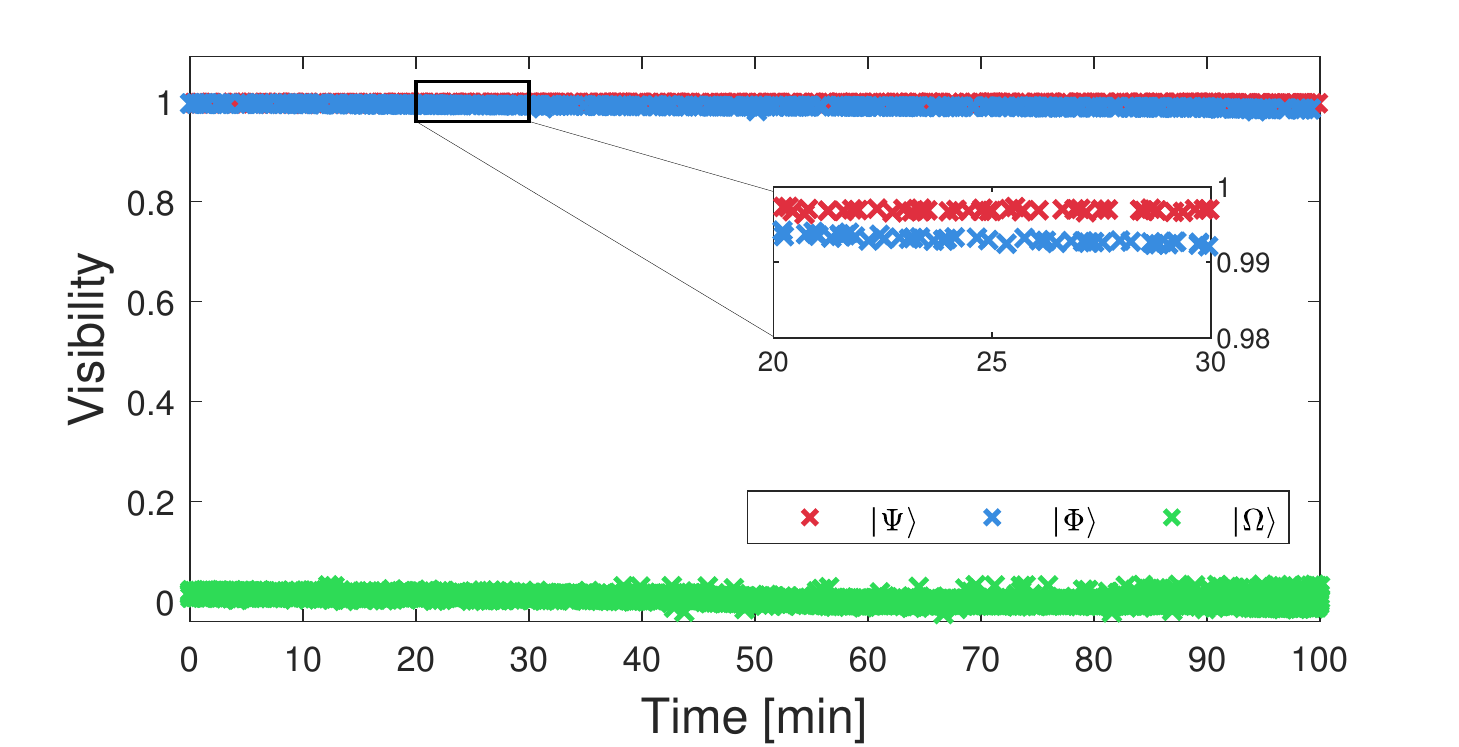}

    \caption{\label{fig:partial_mdi} { \color{black} Achieved visibilities for the states $\ket{\Psi}$ and $\ket{\Phi}$ that are expected to yield deterministic measurement results, together with the visibilities for the balanced state $\ket{\Omega}$ used for randomness generation. } }
\end{figure}

{\color{black}To further demonstrate that the tunability can be used as a building block in measurement-device independent protocols we run a prepare-and-measure type protocol where we use the TBS to prepare either of the orthogonal states $\ket{\Psi}$ and $\ket{\Phi}$ which are expected to yield deterministic measurement results, and a state $\ket{\Omega}$ which yields random (but balanced) measurement outcomes. We prepare the states by applying different voltages to the PM inside the Sagnac for each state. From \cref{fig:countsphasevoltage} we determine the voltages to be  $V_\Psi = 0 \textrm{ V}$,  $V_\Phi = 4.2 \textrm{ V}$ and  $V_\Omega = 2.15 \textrm{ V}$.  For each acquired 32 kilobyte block of data we then randomly choose between preparing either of the states with a probability $\mathrm{Pr}\left\{ \ket{\Psi}\right\}  = \mathrm{Pr}\left\{\ket{\Phi}\right\} = 0.005 $ and $\mathrm{Pr}\left\{\ket{\Omega}\right\} = 0.99$, where $\mathrm{Pr}\left\{ \ket{\bullet}\right\} $ is the probability to prepare a certain quantum state, in order to demonstrate the switching of states necessary to implement measurement-device independent QRNG protocols. In \cref{fig:partial_mdi} we show the recorded visibilities for each of the states as a function of time, where the visibility $\nu_x \in \left[0,1\right]$ for the state $x\in\left\{\ket{\Psi},\ket{\Phi}, \ket{\Omega}\right\}$  is defined $\nu_{x} = \left| \frac{n_\mathrm{early} - n_\mathrm{late}}{n_\mathrm{early} + n_\mathrm{late}} \right| $ where $n_\mathrm{early}$ and $n_\mathrm{late}$ are the detector counts in the early and late bin respectively. We are able to achieve a mean visibility of $\nu_\Psi = 0.997 \pm 3.3\cdot 10^{-4}$,  $\nu_\Phi = 0.990 \pm 3.1\cdot 10^{-3}$ and $\nu_\Omega = 0.004 \cdot 10^{-2} \pm 7.7 \cdot 10^{-3}$ for each of the three states without subtracting dark counts. }

\section{Conclusions}
We have proposed and experimentally demonstrated a photonic quantum random number generator based on a dynamically tunable beamsplitter. The random numbers are generated depending on which output from the beamsplitter the single-photons are detected. By tuning the beamsplitting ratio we are able to maximize the generated entropy of the random sequence. This also helps maximize the entropy in spite of differences in losses following the two outputs. We implemented the tunable beamsplitter through a Sagnac interferometer, in which the tunability is given by a relative phase between the two counter-propagating paths, applied with an electro-optical telecom phase modulator. This technique allows ultra-fast tunability, due to the short response time of the phase modulator. 

We are able to generate a continuous stream of random numbers over 36 hours of continuous operation, showing an average raw generation rate of 131 kbit/s and an average generated entropy of 7.9925 bits per byte in the raw sequence, thus demonstrating a stable source of entropy. {\color{black} As the bitrate is mainly limited by internal delays in the employed driving electronics, by improving the driver circuitry, a considerable bitrate gain can be achieved. By reducing these delays, we would also reduce the lower limits on the time-bin spacing and thus the repetition rate can be increased.} Our setup is robust and built using only off-the-shelf telecom fiber components, thus also showing the practicability of our setup.  Our results can aid in future designs of dynamically tunable quantum random number generators, and serve as a foundation for measurement-device quantum random number generation due to ease of changing the beamsplitting ratio, through the appropriate selection of the applied phase in the Sagnac loop. 

{\color{black}Our system also demonstrates that the tunable beamsplitter enables realization of measurement-device independent quantum random number generators thanks to the high visibility when measuring orthogonal states, as well the fact that the visibilities show minimal degradation over time.}

\section*{Acknowledgments}
We acknowledge financial support from CENIIT Link\"{o}ping University, the Swedish Research Council (VR 2017-04470), QuantERA grant SECRET (VR 2019-00392), the Knut and Alice Wallenberg Foundation through the Wallenberg Center for Quantum Technology (WACQT).

\section*{Author Contributions}
J.A. and A.A. conceived the work together with G.B.X. J.A. developed the electronic control system, and A.A. developed the optical system. Data analysis was done by J.A. with assistance from G.B.X. The work was coordinated by G.B.X. All authors contributed to the writing of the manuscript. 

\section*{Competing Interests}
The authors declare no competing interests.

\section*{References}
\printbibliography[heading=none]

@article{Carine2020,
	title        = {Multi-core fiber integrated multi-port beam splitters for quantum information processing},
	author       = {Jaime Cari{\~{n}}e and Gustavo Ca{\~{n}}as and P. Skrzypczyk and Ivan \v{S}upi\'{c} and N. Guerrero and T. Garcia and L. Pereira and M. A. S. Prosser and G. B. Xavier and A. Delgado and S. P. Walborn and D. Cavalcanti and G. Lima},
	year         = 2020,
	month        = {May},
	journal      = {Optica},
	publisher    = {OSA},
	volume       = 7,
	number       = 5,
	pages        = {542--550},
	doi          = {10.1364/OPTICA.388912},
}

@article{Ma2016,
	title        = {Quantum random number generation},
	author       = {Ma, Xiongfeng and Yuan, Xiao and Cao, Zhu and Qi, Bing and Zhang, Zhen},
	year         = 2016,
	journal      = {npj Quantum Inf.},
	volume       = 2,
	number       = 1,
	pages        = 16021,
	doi          = {10.1038/npjqi.2016.21},
	isbn         = {2056-6387},
	date         = {2016/06/28},
	id           = {Ma2016}
}

@article{Nie2016,
	title        = {Experimental measurement-device-independent quantum random-number generation},
	author       = {Nie, You-Qi and Guan, Jian-Yu and Zhou, Hongyi and Zhang, Qiang and Ma, Xiongfeng and Zhang, Jun and Pan, Jian-Wei},
	year         = 2016,
	month        = {Dec},
	journal      = {Phys. Rev. A},
	publisher    = {American Physical Society},
	volume       = 94,
	pages        = {060301},
	doi          = {10.1103/PhysRevA.94.060301},
	numpages     = 5
}

@article{Ma2011,
	title        = {A high-speed tunable beam splitter for feed-forward photonic quantum information processing},
	author       = {Xiaosong Ma and Stefan Zotter and Nuray Tetik and Angie Qarry and Thomas Jennewein and Anton Zeilinger},
	year         = 2011,
	month        = {Nov},
	journal      = {Opt. Express},
	publisher    = {OSA},
	volume       = 19,
	number       = 23,
	pages        = {22723--22730},
	doi          = {10.1364/OE.19.022723},
}

@article{Naruse2015,
	title        = {Single-photon decision maker},
	author       = {Naruse, Makoto and Berthel, Martin and Drezet, Aur{\'e}lien and Huant, Serge and Aono, Masashi and Hori, Hirokazu and Kim, Song-Ju},
	year         = 2015,
	journal      = {Sci. Rep.},
	volume       = 5,
	number       = 1,
	pages        = 13253,
	doi          = {10.1038/srep13253},
	isbn         = {2045-2322},
	da           = {2015/08/17},
	id           = {Naruse2015},
	ty           = {JOUR}
}

@article{Wang2006,
	title        = {Scheme for a quantum random number generator},
	author       = {Wang,P. X. and Long,G. L. and Li,Y. S.},
	year         = 2006,
	journal      = {J. Appl. Phys.},
	volume       = 100,
	number       = 5,
	pages        = {056107},
	doi          = {10.1063/1.2338830},
	eprint       = {https://doi.org/10.1063/1.2338830}
}

@article{Stefanov2000,
	title        = {Optical quantum random number generator},
	author       = {Andr{\'e} Stefanov and Nicolas Gisin and Olivier Guinnard and Laurent Guinnard and Hugo Zbinden},
	year         = 2000,
	journal      = {J. Mod. Opt.},
	publisher    = {Taylor \& Francis},
	volume       = 47,
	number       = 4,
	pages        = {595--598},
	doi          = {10.1080/09500340008233380},
	eprint       = {https://doi.org/10.1080/09500340008233380}
}

@article{Jennewein2020,
	title        = {A fast and compact quantum random number generator},
	author       = {Jennewein, Thomas and Achleitner,Ulrich and Weihs,Gregor and Weinfurter,Harald and Zeilinger,Anton},
	year         = 2000,
	journal      = {Rev. Sci. Instrum.},
	volume       = 71,
	number       = 4,
	pages        = {1675--1680},
	doi          = {10.1063/1.1150518},
	eprint       = {https://doi.org/10.1063/1.1150518}
}

@book{NielsenBook,
	title        = {Quantum Computation and Quantum Information: 10th Anniversary Edition},
	author       = {Nielsen, Michael and Chuang, Isaac L},
	year         = 2010,
	journal      = {Cambridge University Press, Cambridge},
	publisher    = {Cambridge University Press},
	doi          = {10.1017/CBO9780511976667},
	place        = {Cambridge}
}

@article{Alarcon2020,
	title        = {{Polarization-independent single-photon switch based on a fiber-optical Sagnac interferometer for quantum communication networks}},
	author       = {Alarc{\'{o}}n, Alvaro and Gonz{\'{a}}lez, P and Cari{\~{n}}e, Jaime and Lima, Gustavo and Xavier, Guilherme B},
	year         = 2020,
	journal      = {Opt. Express},
	volume       = 28,
	number       = 22,
	pages        = 33731,
	doi          = {10.1364/oe.408637},
	archiveprefix = {arXiv},
	arxivid      = {2009.02385},
	eprint       = {2009.02385},
}

@article{Herrero-Collantes2017,
	title        = {{Quantum random number generators}},
	author       = {Herrero-Collantes, Miguel and Garcia-Escartin, Juan Carlos},
	year         = 2017,
	journal      = {Rev. Mod. Phys.},
	volume       = 89,
	number       = 1,
	doi          = {10.1103/RevModPhys.89.015004},
	archiveprefix = {arXiv},
	arxivid      = {1604.03304},
	eprint       = {1604.03304}
}

@article{Rukhin2010,
	title        = {{A Statistical Test Suite for Random and Pseudorandom Number Generators for Cryptographic Applications}},
	author       = {Rukhin, Andrew and Soto, Juan and Nechvatal, James},
	year         = 2010,
	journal      = {Nist Special Publication},
	volume       = 22,
	doi          = {10.6028/NIST.SP.800-22r1a},
}

@misc{R.Brown2011,
	title        = {{Dieharder: A random number test suite}},
	author       = {Brown, Robert G.},
	year         = 2011,
	booktitle    = {Dieharder: A random number test suite},
	urldate      = {2021-11-30}
}

@article{Qi2017,
	title        = {{True randomness from an incoherent source}},
	author       = {Qi, Bing},
	year         = 2017,
	month        = {nov},
	journal      = {Rev. Sci. Instrum.},
	publisher    = {American Institute of Physics Inc.},
	volume       = 88,
	number       = 11,
	pages        = 113101,
	doi          = {10.1063/1.4986048},
	archiveprefix = {arXiv},
	arxivid      = {1611.00224},
	eprint       = {1611.00224},
}

@article{Ma2013,
	title        = {{Postprocessing for quantum random-number generators: Entropy evaluation and randomness extraction}},
	author       = {Ma, Xiongfeng and Xu, Feihu and Xu, He and Tan, Xiaoqing and Qi, Bing and Lo, Hoi Kwong},
	year         = 2013,
	journal      = {Phys. Rev. A: At. Mol. Opt. Phys.},
	volume       = 87,
	number       = 6,
	pages        = 62327,
	doi          = {10.1103/PhysRevA.87.062327},
	archiveprefix = {arXiv},
	arxivid      = {1207.1473},
	eprint       = {1207.1473},
}

@article{Micuda2014,
	title        = {{Highly stable polarization independent Mach-Zehnder interferometer}},
	author       = {Mi{\v{c}}uda, Michal and Dol{\'{a}}kov{\'{a}}, Ester and Straka, Ivo and Mikov{\'{a}}, Martina and Du{\v{s}}ek, Miloslav and Fiur{\'{a}}{\v{s}}ek, Jarom{\'{i}}r and Je{\v{z}}ek, Miroslav},
	year         = 2014,
	month        = {aug},
	journal      = {Rev. Sci. Instrum.},
	publisher    = {American Institute of PhysicsAIP},
	volume       = 85,
	number       = 8,
	pages        = {083103},
	doi          = {10.1063/1.4891702},
	archiveprefix = {arXiv},
	arxivid      = {1407.5207},
	eprint       = {1407.5207},
}

@article{Xavier2011,
	title        = {{Stable single-photon interference in a 1 km fiber-optic Mach–Zehnder interferometer with continuous phase adjustment}},
	author       = {Xavier, G. B. and von der Weid, J. P.},
	year         = 2011,
	month        = {may},
	journal      = {Opt. Lett.},
	publisher    = {Optical Society of America},
	volume       = 36,
	number       = 10,
	pages        = 1764,
	doi          = {10.1364/ol.36.001764},
	archiveprefix = {arXiv},
	arxivid      = {1104.2866},
	eprint       = {1104.2866},
}

@article{Reynaud1992,
	title        = {{Interferometric control of fiber lengths for a coherent telescope array}},
	author       = {Reynaud, F and Alleman, J J and Connes, P},
	year         = 1992,
	journal      = {Appl. Opt.},
	volume       = 31,
	number       = 19,
	pages        = 3736,
	doi          = {10.1364/ao.31.003736}
}

@article{Goldwasser1984,
	title        = {{Probabilistic encryption}},
	author       = {Goldwasser, Shafi and Micali, Silvio},
	year         = 1984,
	month        = {apr},
	journal      = {J. Comput. Syst. Sci.},
	publisher    = {Academic Press},
	volume       = 28,
	number       = 2,
	pages        = {270--299},
	doi          = {10.1016/0022-0000(84)90070-9}
}

@article{Hung1997,
	title        = {{The Behavior of the P-Value When the Alternative Hypothesis is True}},
	author       = {Hung, H. M. James and O'Neill, Robert T. and Bauer, Peter and Kohne, Karl},
	year         = 1997,
	month        = {mar},
	journal      = {Biometrics},
	publisher    = {JSTOR},
	volume       = 53,
	number       = 1,
	pages        = 11,
	doi          = {10.2307/2533093},
	issn         = {0006341X},
}

@article{Roberts2018,
	title        = {{Patterning-effect mitigating intensity modulator for secure decoy-state quantum key distribution}},
	author       = {Roberts, G L and Pittaluga, M and Minder, M and Lucamarini, M and Dynes, J F and Yuan, Z L and Shields, A J},
	year         = 2018,
	journal      = {Opt. Lett.},
	volume       = 43,
	number       = 20,
	pages        = 5110,
	doi          = {10.1364/ol.43.005110},
	issn         = {0146-9592},
}

@article{Agnesi2019,
	title        = {{All-fiber self-compensating polarization encoder for quantum key distribution}},
	author       = {Agnesi, Costantino and Avesani, Marco and Stanco, Andrea and Villoresi, Paolo and Vallone, Giuseppe},
	year         = 2019,
	journal      = {Opt. Lett.},
	volume       = 44,
	number       = 10,
	pages        = 2398,
	doi          = {10.1364/ol.44.002398},
	issn         = {0146-9592},
	archiveprefix = {arXiv},
	arxivid      = {1903.00696},
	eprint       = {1903.00696},
}

@article{Avesani2021,
	title        = {{Deployment-ready quantum key distribution over a classical network infrastructure in Padua}},
	author       = {Avesani, Marco and Foletto, Giulio and Padovan, Matteo and Calderaro, Luca and Agnesi, Costantino and Bazzani, Elisa and Berra, Federico and Bertapelle, Tommaso and Picciariello, Francesco and Santagiustina, Francesco and Scalcon, Davide and Scriminich, Alessia and Stanco, Andrea and Vedovato, Francesco and Vallone, Giuseppe and Villoresi, Paolo},
	year         = 2021,
	journal      = {J. Light. Technol.},
	publisher    = {Institute of Electrical and Electronics Engineers Inc.},
	doi          = {10.1109/JLT.2021.3130447},
	issn         = 15582213,
	archiveprefix = {arXiv},
	arxivid      = {2109.13558},
	eprint       = {2109.13558},
}

\end{document}